\documentclass[12pt]{iopart}

\usepackage{graphicx}
\usepackage{fullpage}
\usepackage{latexsym}
\usepackage{amssymb}
\usepackage{wasysym}
\usepackage{color,linegoal}

\newcommand{\nA} {n_\mathrm{A}}

\newcommand{\Kexp} {K_\mathrm{exp}}

\newcommand{\Gammae} {\Gamma_\mathrm{e}}

\newcommand{\phieff} {\Phi_\mathrm{eff}}
\newcommand {\gammadot} {\dot{\gamma}}

\begin{document}

\title{Bulk and interfacial stresses in suspensions of soft and hard colloids}

\author{D. Truzzolillo$^{1}$, V. Roger$^{1}$, C. Dupas$^{1}$, S. Mora$^{2}$, L. Cipelletti$^{1}$}
\address{$1$- Laboratoire Charles Coulomb (L2C), UMR 5221 CNRS-Univ. Montpellier 2, Montpellier, F-France.}
\address{$2$- CNRS, Laboratoire de M\'{e}canique et de G\'{e}nie Civil UMR 5508, F-34095 Montpellier, France}
\ead{domenico.truzzolillo@univ-montp2.fr}


\begin{abstract}

We explore the influence of particle softness and internal structure on both the bulk and interfacial rheological properties of colloidal suspensions. We probe bulk stresses by conventional rheology, by measuring the flow curves, shear stress \textit{vs} strain rate, for suspensions of soft, deformable microgel particles and suspensions of near hard-sphere-like silica particles. A similar behavior is seen for both kind of particles in suspensions at concentrations up to the random close packing volume fraction, in agreement with recent theoretical predictions for sub-micron colloids. Transient interfacial stresses are measured by analyzing the patterns formed by the interface between the suspensions and their own solvent, due to a generalized Saffman-Taylor hydrodynamic instability. At odd with the bulk behavior, we find that microgels and hard particle suspensions exhibit vastly different interfacial stress properties. We propose that this surprising behavior results mainly from the difference in particle internal structure (polymeric network for microgels \textit{vs} compact solid for the silica particles), rather than softness alone.

\end{abstract}

\maketitle

\section{Introduction}

Colloidal suspensions comprise solid particles in the size range 1 nm - 10 $\mu$m suspended in a background fluid. They are ubiquitous in every day life and industrial applications, and they are intensively studied as model systems in condensed matter, \textit{e.g.} for tackling problems such as the glass transition or crystallization. Colloidal suspensions, even in the concentrated regime or in the presence of strong interparticle interactions, are \textit{soft} systems, in that they are significantly deformed or even driven from a solid-like to a fluid-like state by relatively modest external forces. Indeed, how colloidal systems may respond to an external drive is a key property in many applications \cite{Larson}, besides being a rich and fascinating problem \textit{per se}. Accordingly, there has been a great interest in the rheological properties of suspensions, both in the linear regime corresponding to small drives and in the non-linear regime \cite{koumakis2012,Ikeda2012}. The availability of particles in a large variety of shapes \cite{Quan2010,Huang2012,Henzie2011,Rossi2011,Liu2008,Shah2013,Sacanna2013}, with different degrees of softness \cite{koumakis2012,Senff2000,Roovers1993,Truzzolillo2013}, and with  highly tailored interactions (including selective interactions mediated by DNA \cite{Biffi2013} or anisotropic interactions as in patchy colloids \cite{Shah2013}) has further spurred studies on the relationship between the rheological properties of a suspension and the structure and interactions of its constituents.

In particular, the role of particle softness
has been investigated in several recent works, in part thanks to the availability of particles with a controlled degree of softness, \textit{e.g.} solid particles covered by a polymer layer \cite{koumakis2012, Chevigny2011} or microgel particles formed by cross-linked polymers \cite{Senff2000, Senff1999}. There is some consensus that the rheological properties of small, sub-micron hard particles are governed by thermal stresses of entropic origin, which sharply increase well before attaining the random close packing volume fraction, while the flow properties of jammed packings of large, athermal and deformable particles are governed by the elastic energy associated with particle deformation. The behavior of particles at intermediate size and softness scales, such as sub-micron PNiPAM microgels, however, is much more controversial and the nature of the fluid-to-solid transition (glass transition \textit{vs} jamming) in these systems is highly debated \cite{Ikeda2012, Ikeda2013, Nordstrom2010, Yunker2014}.

Soft matter systems may also exhibit intriguing behaviors ruled by \emph{interfacial} stresses, as opposed to the bulk stresses discussed above. For example, recent work has shown that the addition of colloidal particles at the interface between two (immiscible) fluids may modify profoundly the mechanical properties of the interface, by imparting it mechanical rigidity as in ``bijels'' \cite{Stratford2005,Herzig2007} or in non-spherical ``armoured'' bubbles \cite{Subramaniam2005,Abkarian2013}, or by protecting bubbles and drops from coalescence \cite{Binks2006}. Recently, we have explored the role of interfacial stresses in colloidal systems in what is arguably the most minimalist experimental configuration: the sharp interface between a colloidal suspension and its own solvent \cite{Truzzolillo2014}. By investigating the onset of the Saffman-Taylor instability, a hydrodynamic instability ruled by the competition between viscosity and surface tension, in suspensions of PNiPAM microgels exposed to their solvent, we have shown that interfacial stresses develop at the suspension-solvent interface. These stresses will eventually decay at large enough time, because the suspension and the solvent are fully miscible, so that diffusion-driven mixing will finally erase any sharp interface. However, on short enough time scales, these stresses act as an effective interfacial tension, as it was already pointed out for the general case of miscible fluids with composition gradients more than 100 years ago, by mathematician and physicist D. Korteweg.

In \cite{Truzzolillo2014} we have shown that the effective surface tension of microgel suspensions can be rationalized in the framework of Korteweg's theory for miscible fluids. Here, we present new data on the effective surface tension between suspensions of hard, compact silica particles and their own solvent. For both kinds of particles, we also measure the bulk rheological properties, so as to compare the impact of particle structure and softness on the bulk rheology to that on interfacial stress. We find little variation of the bulk rheological properties with particle kind, while microgels and silica particles exhibit vastly different effective surface tension properties. We discuss this surprising result in view of available models for both bulk and interfacial stresses.

\section{Sample characterization}

\subsection{Microgels}\label{microgels}
The soft particles are poly-N-isopropylacrylamide (PNiPAM) microgels synthesized by emulsion polymerization according to the protocol in~\cite{Senff1999} and suspended in water. PNiPAM solutions exhibit a lower critical solution temperature (LCST) close to room temperature, which results in a $T$ dependence of the microgel size. We characterize the $T$-dependent particle size of our microgels by measuring the hydrodynamic radius $R_h$ with conventional dynamic light scattering (DLS)~\cite{Berne1976}, using very diluted suspensions (w/w concentration $c=10^{-5}$). Increasing the temperature leads to a gradual decrease of the particle size: in the temperature range $294.0~\mathrm{K} \le T \le 309.9~\mathrm{K}$, we find that $R_h(T)$ is well approximated by the critical-like function $R_h =\varepsilon(1-T/Tc)^{\beta}$, with $T_c =304.8$ K, $\varepsilon=240.85$ nm and $\beta=0.116$. All the experiments reported in this paper have been performed at $T=293.16$ K, where the microgels diameter is $d=2R_h=330$ nm.

We perform experiments at several particle concentrations, quantified by the effective volume fraction $\Phi_\mathrm{eff}=n_pv$, where $n_p$ is the number density of colloids and $v=\pi d^3/6$ the volume of a single particle at infinite dilution. Note that, because microgels are soft, squeezable particles, the actual particle volume decreases at very high $n_p$, so that samples with $\Phi_\mathrm{eff} > \Phi_{rcp} \approx 0.64$ or even $\Phi_\mathrm{eff} > 1$ may be prepared, where $\Phi_{rcp}$ is the random close packing volume fraction of hard spheres. Experimentally, only the mass fraction $c$ of a suspension can be directly measured, by weighting a small aliquot of the sample before and after removing the solvent by evaporation. Since the microgels are highly swollen, their mass density is essentially the same as that of the solvent; consequently, $c$ and $\phieff$ are proportional, $\Phi_\mathrm{eff} = kc$. We determine the constant $k$ using two independent methods based on the $c$ dependence of the zero-shear viscosity and of the diffusion coefficient, respectively.

\begin{figure}[htbp]
\centering{
\includegraphics[width=8cm]{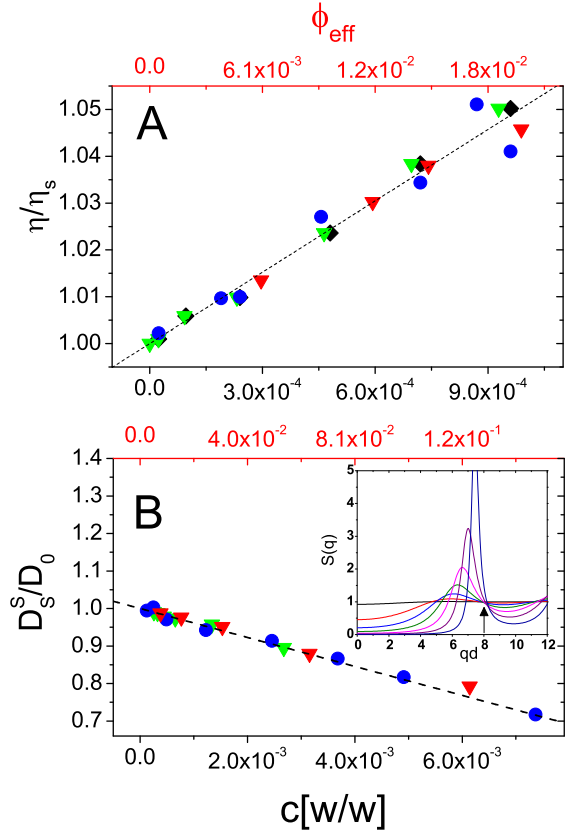}\\}
\caption{Relative viscosity (panel A) and normalized short-time self diffusion coefficient (panel B) of PNiPAM microgel suspensions at different concentrations c (w/w). The various symbols and colors are measurements on suspensions prepared from distinct batches of particles synthesized following the same protocol. The dashed straight lines are linear fits used to extract $k$, as discussed in the text. Inset of B: static structure factor $S(q,d)$ calculated using the Percus-Yevick approximation. All DLS measurement are performed at the $q$ vector indicated by the arrow, where $S(q,d)\approx 1$}
\label{einst-toku}
\end{figure}

Viscosity measurements are performed in the range $0.0023 \le c \le 0.0980$ using an Anton Paar Lovis 2000 ME microviscosimeter. \Fref{einst-toku}A shows that the zero shear viscosity $\eta$ increases linearly with particle concentration, as predicted --for the dilute regime-- by Einstein's formula:
\begin{equation}\label{einstein}
\eta / \eta_0 = 1 + 2.5 \Phi_\mathrm{eff} = 1 + 2.5 k c\,,
\end{equation}
where $\eta_0$ is the viscosity of the solvent. By fitting $\eta / \eta_0$ to a straight line, we determine $k=20.3 \pm 0.6$. Note that the viscosity data are taken for $\Phi_\mathrm{eff} \lesssim 0.02$ (see the upper axis in \fref{einst-toku}A), a regime dilute enough for higher order corrections to (\ref{einstein}) to be negligible.

The short-time self diffusion coefficient of the microgels, $D_S^S$, is obtained by fitting to a simple exponential decay the initial decay of the intermediate scattering function (ISF) measured by DLS, $g_1(\tau) = \exp(-D_S^Sq^2\tau$), where $q$ is the scattering vector~\cite{Berne1976}. As pointed out in~\cite{PuseyJPhysA1978}, the ISF reflects the \textit{self} dynamics only if $g_1$ is measured at a scattering vector such that $S(q,d) = 1$, with $S(q,d)$ the static structure factor. Using the value of $d$ obtained by DLS in the very dilute limit and the Percus-Yevick approximation of $S(q,d)$ for an equivalent hard sphere system~\cite{Hansen}, we fix the scattering angle at $\theta=101~\mathrm{deg}$, corresponding to $q = 24.3~\mu\mathrm{m}^{-1}$, such that $S(q,d)\approx 1$ at all concentrations of interest (see inset of \fref{einst-toku}B). \Fref{einst-toku}B shows that the normalized short-time self diffusion coefficient, $D_S^S/D_0$, decreases linearly with particle concentration (here, $D_0$ is the zero-$c$ extrapolation of $D_S^S$). We model the $c$ dependence of $D_S^S/D_0$ using the Tokuyama-Oppenheim  expression~\cite{Tokuyama1994,Cichocki1999}:
\begin{equation}\label{tokuyama}
D_S^S/D_0 = \frac{1}{1+H(\Phi_\mathrm{eff})} \simeq 1 - 1.83\Phi_\mathrm{eff}  =  1 - 1.83 k c \,,
\end{equation}
where the approximation holds in dilute limit and $H(\Phi_\mathrm{eff})$ is the hydrodynamic function. By fitting $D_S^S(c)/D_0$ to a straight line, we obtain  $k=20.1 \pm 0.4$. This value is fully compatible with that obtained by viscosimetry: the discrepancy between the two methods is less than 1\%, indicating that the combination of these two low-$\Phi_\mathrm{eff}$ methods allows us to determine the absolute volume fraction of our suspensions to within 0.01-0.02 (at most), including at high $\Phi_\mathrm{eff}$. This level of uncertainty is comparable to or better than that typically reported for colloidal suspensions~\cite{poon10}. In the following, $\Phi_\mathrm{eff}$ is calculated using $k = 20.2 \pm 0.5$, obtained by averaging the viscosimetry and DLS values of $k$.

\subsection{Silica Particles}
Silica particles are used as a model system for hard, undeformable particles with short range, nearly hard-sphere-like interactions. The particles are Ludox-TM 50 purchased from Sigma-Aldrich and used without further purification. The particles are charge-stabilized: to partially screen long range electrostatic repulsions, we add a monovalent salt, KCl, adjusting the final salt concentration in all suspensions to 0.05 M. For very dilute suspensions, the hydrodynamic diameter measured by DLS is $d=2R_h=36$ nm, the same value as for salt-free suspensions. This suggests that the salt concentration is low enough to avoid particle aggregation, which may occur due to van der Waals interactions when electrostatic repulsions are only partially screened. Since rheology and surface tension measurements are performed also at high concentration, we further monitor the suspension stability in the concentrated regime ($c=0.545$ w/w, corresponding to $\Phi_\mathrm{eff} = 0.558$ as discussed below). We measure $\tau_\alpha$, the relaxation time of the suspension at $qd = 0.80$, by means of multispeckle DLS, using the apparatus described in \cite{ElMasriJPCM2005}, for waiting times ranging from one day up to one week after preparing the suspension. We find $\tau_\alpha=4.4 \pm 0.1$ s independently of sample age, thus confirming that no significant aggregation occurs up to time scales larger that those of the experiments discussed here.

\begin{figure}[htbp]
\centering{
\includegraphics[width=8cm]{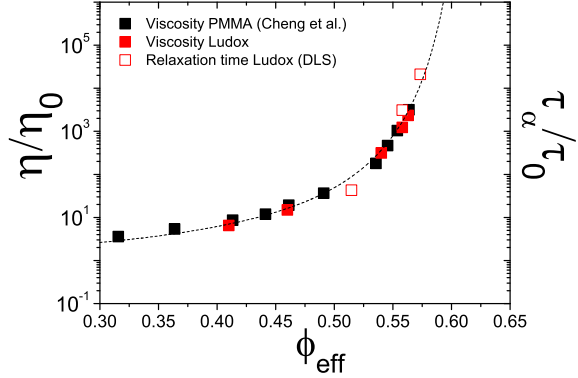}\\}
\caption{Relative viscosity (red full squares) and normalized microscopic relaxation time measured by DLS (red empty squares) of Ludox suspensions as a function of the effective volume fraction $\Phi_\mathrm{eff}$. Reference hard-sphere data for the relative viscosity have been taken from \cite{Chen2002} (black full squares) and fitted by the Doolittle equation (dotted line).}
\label{ViscosityPhi}
\end{figure}

As for the microgels, only the mass fraction $c$ of the silica particle suspensions can be directly measured. The two methods used to determine $\phieff$ and discussed above, however, cannot be applied to the silica particles, because electrostatic repulsions are not fully screened. Indeed, Einstein's law (\ref{einstein}) holds for particles interacting solely via excluded volume interactions, while the hydrodynamic function $H(\phieff)$ in (\ref{tokuyama}) 
cannot be determined without a detailed knowledge of the actual interparticle potential, which is experimentally difficult to achieve. We thus use a different approach and determine $\phieff$ by matching the concentration dependence of the viscosity in concentrated Ludox suspensions ($0.4<\Phi_\mathrm{eff}<0.56$) to that of a reference PMMA hard sphere system \cite{Chen2002}. For solid, compact particles, one expects the effective volume fraction to be related to the mass fraction $c$ by $\phieff=k'c\rho_s/[\rho_p-c(\rho_p-\rho_s)]$, where $\rho_s$ and $\rho_p$ are the solvent and particle mass density, respectively. The factor $k'$ accounts for deviations with respect to hard-sphere behavior, due to particle interactions: for hard spheres, $\phieff$ coincides with the ``geometrical'' volume fraction $n_p \pi d^3/6$, corresponding to $k'=1$, while in the presence of repulsive interactions we expect the effective particle volume to be larger, yielding $k'>1$. \Fref{ViscosityPhi} shows the result of the mapping obtained from the known values of the particle and solvent density, $\rho_p=2.2$ g/ml and $\rho_s=1$ g/ml, and using $k' = 1.58$, corresponding to a particle ``interaction diameter'' $d\sqrt[3]{k'}= 42~\mathrm{nm}$, about 17\% larger than the hydrodynamic diameter in the limit of infinite dilution. With this choice of $k'$, a very good agreement is observed between the viscosity of hard spheres and that of the Ludox suspensions over two decades in $\eta$. To further test the robustness of this procedure, we measure the $\phieff$ dependence of $\tau_\alpha$. 
\Fref{ViscosityPhi} shows that the concentration dependence of the microscopic relaxation time normalized by its $c \rightarrow0$ value is in good agreement with that of the relative viscosity, which we extend to a broader range of $\phieff$ using a Doolittle fit \cite{Chen2002},
$\eta/\eta_0=\exp[\Lambda\phi_\mathrm{eff}/(\phi_\mathrm{eff}-\phi_{rcp})]$ (dotted line in \Fref{ViscosityPhi}). Although the exact relation between $\tau_\alpha$ and $\eta$ is still debated (see e.g.~\cite{poon10}), no strong deviations from a simple $\tau_\alpha \sim \eta$ scaling are expected in the range of $\phieff$ probed here. Thus, the agreement between the trend of these two quantities as a function of effective volume fraction confirms the soundness of the procedure used to determine $\phieff$.


\section{Bulk rheology}\label{sec:bulkrheo}

\begin{figure}[htbp]
\centering{
\includegraphics[width=8cm]{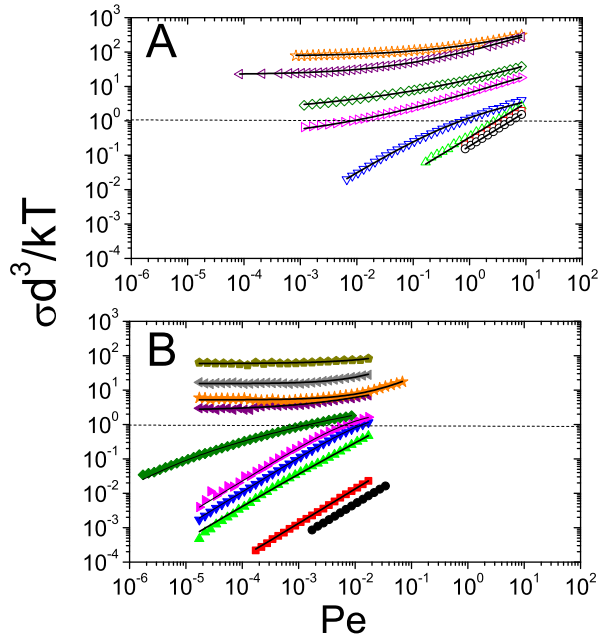}\\}
\caption{Flow curves: stress versus shear rate under steady shear for samples with different effective volume
fractions. MICROGELS (Panel A): from bottom to top, $\Phi_\mathrm{eff}=0.20$, $0.27$, $0.33$, $0.46$, $0.55$, $0.61$, $0.92$, and $1.2$.
LUDOX PARTICLES (Panel B): from bottom to top, $\Phi_\mathrm{eff}=0.41$, $0.46$, $0.540$, $0.558$, $0.563$, $0.573$, $0.591$, $0.607$, $0.621$, and $0.632$.}
\label{fig:rheo-flowcurves}
\end{figure}

\Fref{fig:rheo-flowcurves} shows the flow curves, shear stress $\sigma$ \textit{vs} shear rate $\gammadot$, for microgels and Ludox suspensions at various $\phieff$ (panel A and B, respectively). The flow curves are obtained by performing steady rate rheology experiments, using a cone-plate geometry (cone diameter = 50 mm, cone angle = 0.0198 rad), except for microgel suspensions in the range $0.4<\Phi_\mathrm{eff}\le 1.2$, for which a 25 mm-plate with a roughened surface has been used to avoid wall slip. The flow curves have been measured both by increasing sequentially the shear rate and by decreasing it, starting from its largest value. No difference were observed depending on the chosen protocol. In order to allow for a comparison with previous works, the data are presented in reduced stress and shear rate units, the former being normalized by $\sigma_T = k_bT/d^3$, the typical stress created by thermal fluctuations, while the shear rate is expressed in units of the (bare) Peclet number $Pe = \gammadot d^2/D_0$, where $d^2/D_0$ is the characteristic time for a particle to diffuse over a distance equal to its diameter, in the $\phieff \rightarrow 0$ regime.

At low volume fractions, for both hard and soft particles $\eta \sim \gammadot$ throughout the whole range of shear rates investigated, indicating Newtonian behavior, as usually observed in simple fluids. As $\phieff$ increases, increasingly stronger deviations with respect to Newtonian flow are observed, until yield stress behavior is clearly seen, for $\Phi_\mathrm{eff}>0.61$ in microgel suspensions and for $\Phi_\mathrm{eff}>0.59$ in Ludox suspensions. Such behavior is quite general; it has predicted in recent models of complex fluids \cite{Ikeda2013,Basu2014} and has been observed experimentally for a wide class of suspensions comprising both soft and hard particles \cite{koumakis2012,Basu2014,petekidis2009}. To better characterize the flow behavior, we fit the flow curves employing various functional forms depending on $\phieff$. The fits will also be used in \sref{sec:fingers_experiments} to obtain by extrapolation the shear-dependent viscosity in the $\gammadot$ regime relevant to the Hele-Shaw experiments.

For Newtonian fluids, the fitting function is simply $\sigma(\gammadot) = \eta \gammadot$. For samples showing weak shear thinning, the flow curves have been fitted using a Cross-like equation: $\sigma(\dot{\gamma})=\frac{\eta_{0,s} \dot{\gamma}}{1+(P\dot{\gamma})^m}$, where $\eta_{0,s}$ is the zero-shear viscosity of the suspension, $1/P$ is the characteristic shear rate denoting the onset of the shear thinning and $m$ the shear thinning exponent \cite{Mewis}. For intermediate concentrations, right below the onset of a dynamical yield stress, the shear stress is well described by a linear combination of two power laws: $\sigma(\dot{\gamma})=G\dot{\gamma}^g+L\dot{\gamma}^l$. This functional form has been successfully used to describe the flow of glassy star polymers in good solvent conditions~\cite{Erwin2010}, a model for ultra soft particles. Finally, for suspensions showing a yield stress behavior we use the Herschel-Bulkley equation~\cite{Cloitre2003}: $\sigma(\dot{\gamma})=\sigma_Y+\lambda\dot{ \gamma}^{\beta}$, where $\sigma_Y$ is the yield stress. The numerical values of the fitting parameters for all curves shown in \fref{fig:rheo-flowcurves} are reported as supplementary information \cite{SuppInfo}.

\begin{figure}[htbp]
\centering{
\includegraphics[width=8cm]{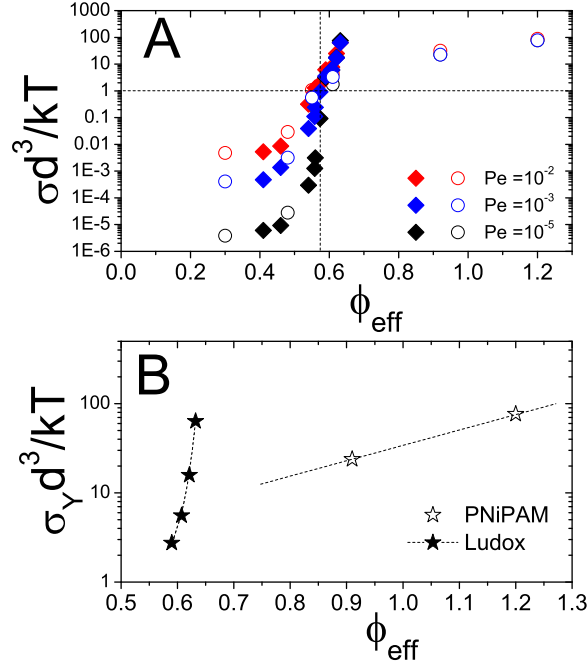}\\}
\caption{A: Normalized shear stress as a function of $\Phi_\mathrm{eff}$ for three different Peclet numbers in the thermal sector ($Pe<1$), for microgels (empty symbols) and solid particles (solid symbols). Some of the values reported here have been obtained by extrapolation using the fitting functions described in the main text. B: $\phieff$ dependence of the normalized yield stress for microgels and Ludox suspensions obtained via Herschel-Bulkley fits.}
\label{fig:rheo-yield}
\end{figure}

Recently, there has been a lively debate on the nature of the fluid-to-solid transition in amorphous solids \cite{koumakis2012,Ikeda2012,Ikeda2013,Nordstrom2010,Yunker2014,petekidis2009}. In the most general scenario proposed in \cite{Ikeda2012,Ikeda2013}, up to two distinct transitions may be observed in thermal systems comprising soft particles: upon increasing $\Phi$, a glass transition is first observed at $\Phi<\Phi_{rcp}$, signalled by the emergence of a yield-stress plateau at low $\gammadot$. The stress scale of this plateau is that of thermal stress, $k_BT/d^3$. Deformable particles can be further concentrated at volume fractions beyond $\Phi_{rcp}$, where the yield stress $\sigma_Y$ steeply increases to higher stress values, ruled by the energetic cost to deform a particle. Experimentally, it is typically difficult to observe these two transitions in the same system, because the range of accessible $\gammadot$ may be too small, or because the particle size and softness may be such that the two plateaus are not well distinct. Indeed, in most cases only one transition is seen; as a consequence, contrasting interpretations have been given to the flow curves of dense colloidal suspensions, especially in the case of soft particles such as PNiPAM microgels  \cite{Ikeda2013,Nordstrom2010,Yunker2014,petekidis2009}, where both mechanism are a priori possible. As pointed out in \cite{Ikeda2012,Ikeda2013}, it is then important to examine the numerical values of the stress and $\gammadot$ scales associated with the emergence of yield stress behavior. The data of \fref{fig:rheo-flowcurves} belong essentially to the ``thermal sector'' defined in \cite{Ikeda2012,Ikeda2013}, where $Pe \lesssim 1$ and $\sigma_Y$ is of the order of the thermal stress $\sigma_T$. This suggest that the fluid to solid transition observed here is ruled by glassy behavior, not only for the Ludox hard particles, but also for the deformable microgels, in agreement with \cite{petekidis2009}.

To compare in more detail the behavior of the two kinds of suspensions, we plot in \fref{fig:rheo-yield}A the $\phieff$ dependence of $\sigma$ for three values of the Peclet number, in the regime $Pe<<1$. Two regimes can be distinguished. For $\sigma d^3/k_bT \le 1$  the stress shows almost no dependence on the softness of particles: the normalized stress for silica and microgels suspensions collapses on the same curve, an unambiguous evidence that both are dominated by thermal fluctuations. For $\sigma d^3/k_bT>1$, the athermal contribution to the overall stress starts emerging: The energy scale dictated by the interparticle interactions contributes to energy dissipation, producing larger stresses for the harder particles, i.e. the Ludox. Note that, because our silica particles are close to hard spheres, for the Ludox suspensions this regime extends only over a very limited range of $\phieff$, close to and below $\Phi_{rcp}$. In the region where $\sigma > \sigma_T$, a qualitative difference in terms of internal dynamics and bulk rheology must be expected for particles interacting via different kind of potentials. To underline this last point, \fref{fig:rheo-yield}B shows the yield stress extrapolated via the Herschel-Bulkley fits for the two sets of particles in the very high $\phieff$ regime. The yield stress of Ludox suspensions rapidly increases as a function of $\phieff$ on approaching $\Phi_{rcp}$. For the microgel suspensions, similar values of $\sigma_Y$ are attained, but the growth is much gentler, owing to particle deformability.

We end this section on the bulk rheology by summarizing the main results: no significant differences are observed in the behavior of the two suspensions up to packing fractions close to $\Phi_{rcp}$, because for both systems the particles are small enough for thermal stresses to fully dominate their mechanical behavior. Only at larger $\phieff$, where particle contacts are ubiquitous, does the difference in softness play a major role, as shown by the sharp growth of $\sigma_Y(\phieff)$ for the Ludox suspensions, to be contrasted to the smooth increase observed for microgels.

\section{Viscous fingering: overview of theory}

In order to measure the (transient) surface tension between the colloidal suspensions and their own solvent, we analyze the patterns formed when pushing the (less viscous) solvent in the (more viscous) suspensions. As shown in \cite{Saffman1958} for immiscible fluids, in the Hele-Shaw geometry where the fluids are confined in the thin gap between two parallel plates the interface between the displacing medium and the displaced fluid becomes instable and develops distinctive finger-like fluctuations. Crucially to our experiments, the number of fingers in this Saffman-Taylor instability is directly related to the (effective) interfacial tension. In this section, we briefly review the formalism required to retrieve the interfacial tension from the observed patterns. In the following, we will assume that no mixing occurs during the interface propagation, which is of course the case for immiscible fluids, but which also applies  to miscible fluids on short enough time scales, as in our experiments.

We focus on a radial geometry, where the more viscous fluid initially occupies a disk-like volume between the plates and the displacing fluid is injected through a hole in one of the plates, centered with respect to the first fluid. No lateral confinement is imposed to the fluids. At the beginning of the injection, the interface between the two fluids is stable and has a circular shape, when observed from above the two plates. As the instability develops, deviations from a circular shape are seen, which appear as ``fingers'' whose number and  length increase with time, see figure 5a)-c). The early stages of the instability are conveniently described by decomposing in Fourier modes the fluctuations of the interface around the unperturbed, perfectly circular shape. The experimental observable is the number of fingers, which has to be related to the dominant modes in the Fourier analysis. In the last 50 years, many authors have associated the number of fingers to the Fourier mode with fastest growth rate \cite{Paterson1981,Paterson1985,Bataille1968,Wilson1975,Miranda1998,Alvarez-Lacalle2004}. This analysis is particularly convenient computationally, because an analytical expression can be derived for $n_f(t)$, the order of the mode that grows faster. However, it was pointed out in \cite{Miranda1998} that in radial Hele-Shaw experiments each Fourier mode evolves at a different time-dependent rate. Thus, the fastest-growing mode at any given time does not coincide in general with the mode that has developed the maximum amplitude.

In recent theoretical and experimental studies \cite{Truzzolillo2014,Dias2013} it has been proposed that the number of fingers observed at a given time is well approximated by the order of the mode with maximum amplitude at that time. In the following, we provide numerical evidence to support this assumption and calculate an asymptotic expression for the maximum-amplitude mode from which the interface tension may be obtained. We start from the analysis of the Saffman-Taylor instability by Miranda and Widom~\cite{Miranda1998}, where the perturbation around a circular interface due to the instability is decomposed in Fourier modes of (complex) amplitude $\zeta_n(t)$. Assuming that the noise giving rise to the instability is a complex number $\zeta_n^0$, with a random phase and a $n$-independent modulus, the time-dependent amplitude of the $n$-th mode can be written as~\cite{Miranda1998}:
\begin{equation}\label{eq:zeta}
\fl\zeta_n(t)=\zeta_n^0\left\{\left[K(t)\frac{(nA-1)}{n^2(n-1)}^{nA-1}\right]\exp\left[(nA-1)\left(\frac{1}{K(t)}\frac{n(n^2-1)}{nA-1}-1\right)\right]\right\} \,.
\end{equation}
Here, $A=(\eta_2-\eta_1)/(\eta_2+\eta_1)>0$ is the viscosity contrast between the two fluids, $K(t)=[r(t)Q]/(2\pi\beta)$, where $r(t)$ is the distance from the center of the cell of the unperturbed fluid-fluid interface, $Q$ is the area invaded by the injected fluid per unit time, and $\beta=b^2\Gamma/[12(\eta_1+\eta_2)]$, with $b$ the cell gap and $\Gamma$ the interfacial tension between the two fluids. Note that (\ref{eq:zeta}) only holds for $nA >1$. In our experiments, the viscosity contrast is sufficiently large for this inequality to be fulfilled for $n \ge 2$.

\begin{figure}[htbp]
\centering{
\includegraphics[width=12cm]{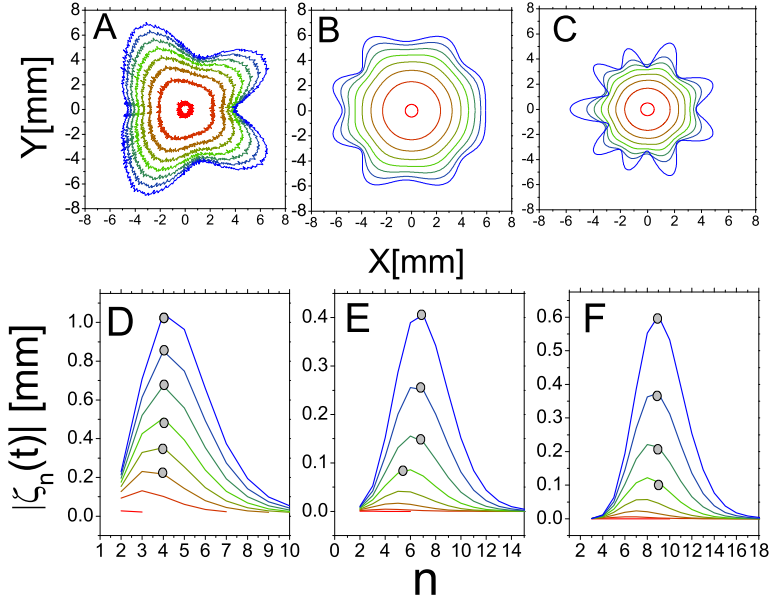}\\}
\caption{Theoretical time evolutions of the fingering patterns (top row) and correspondent Fourier mode amplitudes (bottom row). The time between successive data sets is 0.1 s. The dots in the bottom row plots indicate the number of fingers visible in the patterns in the top row. The interface patterns are generated using (\ref{eq:zeta}), with the following sets of parameters:
A,D: $\Gamma=4$ $mN/m$, $\dot{V}=8\cdot10^{-9}$ $m^3/s$, $\eta_1=10^{-3}$ $Pa\cdot s$, $\eta_2=1$  $Pa\cdot s$, $|\zeta_n(0)|=10^{-8}$ $m$. B,E: $\Gamma=4$ $mN/m$, $\dot{V}=8\cdot10^{-9}$ $m^3/s$, $\eta_1=10^{-3}$ $Pa\cdot s$, $\eta_2=4$  $Pa\cdot s$, $|\zeta_n(0)|=10^{-9}$ $m$. C,F: $\Gamma=4$ $mN/m$, $\dot{V}=8\cdot10^{-9}$ $m^3/s$, $\eta_1=10^{-3}$ $Pa\cdot s$, $\eta_2=6$  $Pa\cdot s$, $|\zeta_n(0)|=10^{-9}$ $m$.}
\label{fig:max-ampl}
\end{figure}

We show in the top row of \fref{fig:max-ampl} a top view of the theoretical interfaces calculated using equation (\ref{eq:zeta}) for three sets of realistic fluid and geometrical parameters. In each panel, the interface is plotted at several times after starting the injection of the less viscous fluid, from $t= 10^{-4}$ s in steps of $0.1$ s. The bottom row shows the corresponding temporal evolution of the amplitude of the Fourier modes, as a function of the mode number. The dots in the bottom-row plots indicate the number of fingers observed in the patterns shown in the top row. In the very early stages, the amplitude of the instability is too small for the number of fingers to be reliably determined; accordingly, no dots are assigned to the corresponding curves $\zeta_n(t)$. As soon as the fingers become visible, their number tracks very well the order of the mode with maximum amplitude, as shown by the fact that in most cases the dots correspond to the peak of the mode amplitude distributions shown in \fref{fig:max-ampl}D-F, the maximum deviation being $|\Delta n| = 1$. This numerical analysis demonstrates that the number of fingers essentially coincides with $n_A$, the order of the mode with maximum amplitude. To make further progress, we need an expression relating $n_A$ to the geometrical and fluid parameters, including the interfacial tension. Since an analytical expression for $n_A$ cannot be found, we derive in the following a useful expression for this mode in the limit $n_A>>1$.

We calculate $n_A$ by solving $d\zeta_n(t)/dn=0$. Using (\ref{eq:zeta}), one finds that  $n_A$ is the solution to
\begin{eqnarray}\label{zeta1}
&\zeta_n^0\left\{\left(K\frac{(nA-1)}{n^2(n-1)}^{nA-1}\right)\exp\left[(nA-1)\left(\frac{1}{K}\frac{n(n^2-1)}{nA-1}-1\right)\right]\right\}\times \nonumber\\
 &\left\{A(\frac{n(n^2-1)}{K(nA-1)}-1)+\frac{1}{K}\left[\frac{(nA-1)(3n^2-1)-nA(n^2-1)}{(nA-1)} \right]\right.+\\ &+\left.n\left(\frac{A}{n}-\frac{2(nA-1)}{(n^2-1)}-\frac{(nA-1)}{n^2}\right)+A\ln\left[\frac{K(nA-1)}{n(n^2-1)}\right]\right\}=0 \nonumber
\end{eqnarray}

The first factor in curly brackets is non-zero for any $n \ge 1$. Hence, (\ref{zeta1}) is satisfied only if the second factor in curly brackets vanishes, which, in the asymptotic limit $n>>1$, yields
\begin{equation}\label{zeta2}
\frac{3n^2}{K}-3A+A\ln\left(\frac{KA}{n^2}\right)=0\,.
\end{equation}
Equation~(\ref{zeta2}) has two real solutions: $n_{A,1}=\sqrt{KA}$ and $n_{A,2}=\sqrt{-W\left(-3e^{-3}\right)KA/3}$, where $W(x)$ is the Lambert function satisfying $x=W(x)e^{W(x)}$. Note that $n_{A,1} > n_f$ while $n_{A,2} < n_f$, where
\begin{equation} \label{eq:nf}
n_f=\sqrt{KA/3}
\end{equation}
is the mode with the maximum growth rate as obtained in~\cite{Miranda1998}. A numerical analysis of the problem shows that the number of fingers grows with time, as confirmed by the experiments and the simulated interfaces (see e.g. \fref{fig:max-ampl}B and E). Thus, at any time the mode with maximum growth rate must be larger than that with maximum amplitude. It follows that the first solution, $n=n_{\mathrm{A,1}}> n_f$, is non-physical. The final expression for the mode with the maximum amplitude is then
\begin{equation}\label{eq:nA}
n_A=\alpha n_f \,,
\end{equation}
with $\alpha=\sqrt{-W\left(-3e^{-3}\right)} \approx 0.422$. Together with (\ref{eq:nf}) and the definitions of $K$ and $A$ (see (\ref{eq:zeta})), (\ref{eq:nA}) provides the link before the experimentally observed number of fingers and the interface tension.

Before recasting this expression in a way that is more suitable to analyze the experiments, two remarks are in order. First, although equation~(\ref{eq:nA}) has been derived in the limit $n>>1$, we have shown \cite{Truzzolillo2014} that, for realistic choices of the fluid and geometrical parameters, equation~(\ref{eq:nA}) represents an excellent approximation to the full solution already for $n_A \geqslant 2$. Thus, in the next section we will identify the number of fingers observed in experiments with $n_A$ as calculated from equation (\ref{eq:nA}). Second, in the above discussion we have implicitly assumed that the viscosity contrast is independent of the rate at which the invading fluid is injected, \textit{i.e.} that both fluids are Newtonian. This is clearly not the case for our colloidal suspensions at large $\phieff$, as seen in \fref{fig:rheo-flowcurves} (recall that $\eta(\dot{\gamma}) = \sigma/\dot{\gamma}$). To take into account the shear-thinning behavior of the suspensions, we use the shear-rate-dependent viscosity $\eta_2(\dot{\gamma_r})$ for evaluating the viscosity contrast $A$, where $\dot{\gamma_r}$ is the shear rate at the position of the interface. This choice relies on the assumption that the wavelength of the perturbation at its onset is not drastically changed by the non-Newtonian features of the suspension. Numerical work on the Saffman-Taylor instability in a radial Hele-Shaw geometry supports this scenario \cite{Sader1994,Kondic1998}, by showing that the non-Newtonian character of the fluids does not change qualitatively the instability, but just accelerates (resp., delays) its onset for shear-thinning (resp., shear thickening) fluids. This choice is also supported by previous works \cite{Bonn2005,Lindner2000} on the Hele-Shaw instability between immiscible non-Newtonian fluids in a rectangular geometry, where the dynamics of the fingers was described by a generalized Darcy law where the Newtonian viscosity was replaced by the shear rate-dependent viscosity.

Having identified the viscosity of the more viscous fluid with the shear-rate-dependent viscosity of the colloidal suspensions, it is convenient to introduce a ``finger function'' \cite{Truzzolillo2014} that, using (\ref{eq:nf}), (\ref{eq:nA}) and the expressions for the fluid and geometrical parameters given after (\ref{eq:zeta}), is shown to be proportional to the shear rate at the injection hole, $\dot{\gamma}_I$, with a proportionality coefficient that directly yields the desired effective interfacial tension:
\begin{equation}\label{eq:Kappaexp}
\Kexp^* \equiv \frac{b}{r}\frac{\left[\frac{3\nA^2}{\alpha^2}-1\right]} {\left[4r_0(\eta_2(\dot{\gamma}_r)-\eta_1)\right]}=\frac{1}{\Gamma_e}\dot{\gamma}_I \,.
\end{equation}
In (\ref{eq:Kappaexp}) all quantities but $\Gammae$ are known or experimentally measurable: $\dot{\gamma}_I=3Q(2\pi r_0 b)^{-1}$ (assuming Poiseuille flow), where $r_0$ is the radius of the injection hole and $Qb$ the imposed flow rate; $n_A$ is obtained by counting the fingers at the onset of the instability; $r = \sqrt{Qbt/\pi}$ is the radius of the unperturbed interface at time $t$; $\eta_1$ is the viscosity of the solvent, and $\eta_2(\dot{\gamma}_r) = \sigma_{fit}/\dot{\gamma}_r$, where $\sigma_{fit}$ is the fit to the flow curve discussed in \sref{sec:bulkrheo}. Finally, the shear rate at the position $r$ of the interface is evaluated as  $\dot{\gamma}_r =4r_0\dot{\gamma}_I/r$, again assuming Poiseuille flow. \Eref{eq:Kappaexp} has been shown to quantitatively capture the behavior of immiscible, Newtonian fluids~\cite{Truzzolillo2014}. It will be used in the next section to measure the effective interfacial tension between our colloidal suspensions and their own solvent.

\section{Off-equilibrium interfacial tension: experiments}
\label{sec:fingers_experiments}

We investigate the $\phieff$ dependence of $\Gammae$ for both microgels and Ludox suspensions by imaging the viscous fingers in Hele-Shaw experiments. For a given $\phieff$, we perform experiments at various $\dot{\gamma}_\mathrm{I}$, always keeping the injection rate high enough for diffusion-driven mixing between the injected solvent and the suspension to be negligible. The Hele-Shaw cell consists of two square glass plates of side $L=25$ mm separated by spacers fixing the gap at $b=0.5$ mm. The cell is first filled with the suspension to be studied. The less viscous fluid (the same solvent as that of the suspension, but died with 0.5\% w/w of methylene blue for visualization purposes) is then injected through a hole in the center of the top plate ($r_0=0.5$ mm). The viscosity of the solvent is $\eta_1 = 1.011~\mathrm{mPa~s}$. The injected volume per unit time, $Qb$, is controlled via a syringe pump. Temperature is fixed at $T = 293\pm 0.1~\mathrm{K}$ by means of a Peltier element, with a circular hole of radius $8.5$ mm for optical observation. A fast CMOS camera (Phantom v7.3 by Vision Research) run at 100 to 3000 frames $\mathrm{s}^{-1}$ is used to image the sample during injection.

\begin{figure}[htbp]
\centering{
\includegraphics[width=8cm]{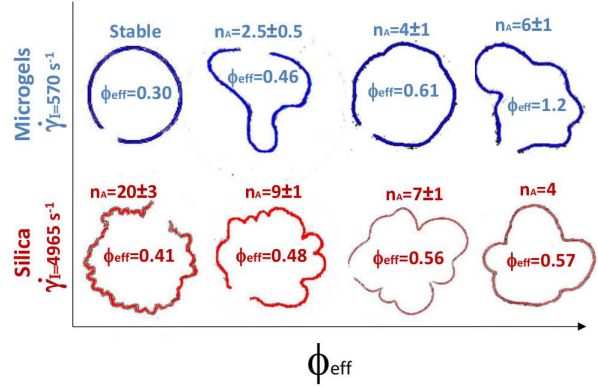}\\}
\caption{Solvent-suspension interface for PNiPAM microgels and Ludox suspensions, for various effective volume fractions and shear rate at the injection point as indicated by the labels.}
\label{fig:Patterns-exp}
\end{figure}

Typical images of the interface between the two fluids are shown in \fref{fig:Patterns-exp}, where the instabilities are visible for both water/microgel and water/Ludox suspensions.
Further images of the two-fluid interface instability at different shear rates are reported in \cite{SuppInfo}.\\
The number $n_A$ of fingers has been evaluated by counting the number of flex points along the interface. Such number is twice the number of the fingers. To improve the accuracy of the estimate of $n_A$, we measure it for two nearby positions of the interface: at the experimentally observable onset of the instability (corresponding to an average distance $r'$ of the interface from the center of the cell) and when the average radius has grown by $\Delta r = 1$ mm. The number of fingers is then calculated as $n_A=[n_A(r')+ n_A(r'+\Delta r)]/2$; the corresponding value of the average radius used in the experimental finger function (\ref{eq:Kappaexp}) is $r = r' + \Delta r /2$.  Finally, both $n_A$ and $r$ thus obtained are further averaged over 2 or 3 experiments performed in the same conditions (same $\phieff$ and injection rate).

\begin{figure}[htbp]
\centering{
\includegraphics[width=8cm]{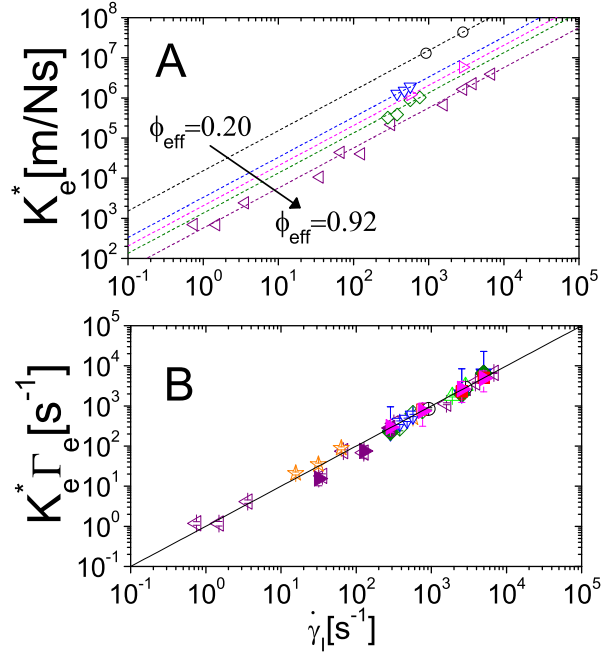}\\}
\caption{A: Finger function $\Kexp^*$ defined in (\ref{eq:Kappaexp}) \textit{vs} the shear rate at the injection hole, $\gamma_{I}$, for microgels suspensions with effective volume fraction in the range $0.2<\Phi_\mathrm{eff}<0.92$ (same symbols as in \fref{fig:rheo-yield}). The dashed lines are linear fits. B: Scaled finger function $\Gamma_e \Kexp^*$ for microgel (open symbols) and Ludox (solid symbols) suspensions at various $\phieff$ (same symbols as in \fref{fig:rheo-yield}). The dashed line is the master curve $\Gamma_e \Kexp^*=\dot{\gamma}_{I}$ predicted by (\ref{eq:Kappaexp}).
}
\label{fig:Ke}
\end{figure}

\Fref{fig:Ke}A shows the finger function $K_e^*$ \textit{vs} the shear rate at the injection hole for microgel suspensions at selected volume fractions in the range $0.2<\Phi_\mathrm{eff}<0.92$. The dashed lines are linear fits to the data, showing that (\ref{eq:Kappaexp}) captures very well the behavior of both Newtonian and non-Newtonian suspensions. We emphasize that strong deviations from the linear scaling $\Kexp^* \sim \dot{\gamma}_I$ would be observed for the high-$\phieff$ suspensions if the zero-shear viscosity was used to evaluate the viscosity contrast, rather than the shear-rate-dependent viscosity. Thus, \fref{fig:Ke}A validates \textit{a posteriori} the choice made in writing (\ref{eq:Kappaexp}). For each $\phieff$, the effective interfacial tension is obtained from a linear fit of $\Kexp^* (\dot{\gamma}_I)$. \Fref{fig:Ke}B demonstrates that very good fits are obtained for both microgel and Ludox suspensions, by showing that the law $\Kexp^* \Gammae = \dot{\gamma}_I$ predicted by (\ref{eq:Kappaexp}) is very well verified over 4 decades.

\begin{figure}[htbp]
\centering{
\includegraphics[width=8cm]{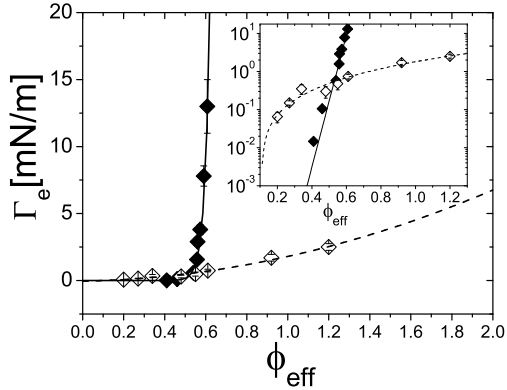}\\}
\caption{Effective interfacial tension $\Gammae$
between colloidal suspensions and their own solvent as a function of colloid effective volume fraction. Open symbols: microgels (data from \cite{Truzzolillo2014}); solid symbols: Ludox suspensions. The dashed (solid) lines are quadratic (exponential) fits, as discussed in the text. Inset: same data plotted in semilogarithmic scale.}
\label{fig:Gamma-phi}
\end{figure}

The effective interfacial tension obtained from the fits shown in \fref{fig:Ke} is plotted as a function of $\phieff$  in \fref{fig:Gamma-phi}. The behavior of $\Gammae$ depends dramatically on the kind of particles: for microgels, $\Gammae$ increases smoothly and mildly with increasing volume fraction. For the Ludox suspensions, by contrast, the effective surface tension is smaller than for the microgels at low $\phieff$ (see inset), but it increases sharply as $\phieff$ approaches the random close packing fraction, resulting in a growth of $\Gammae$ by more than three decades over the range of packing fractions explored in our experiments. Such different behavior is also reflected in the opposite trends seen in the interface patterns, \fref{fig:Patterns-exp}: while for microgels the number of fingers increases with $\phieff$, indicating that the destabilizing effect of the growing viscosity contrast prevails on the stabilizing effect of interfacial tension, for the Ludox suspension $n_A$ decreases with packing fraction, suggesting that a massive growth of $\Gammae$ must more than offset the increase of the viscosity contrast.

The behavior of the microgels has been rationalized in \cite{Truzzolillo2014} in terms of the so-called Korteweg, or square-gradient, model. We briefly recall here the main features of the model. As it was already recognized by Korteweg in 1901~\cite{Korteweg1901}, on time scales shorter than that of interface relaxation due to diffusive mixing, interfacial stresses between miscible fluids exist, whose effect is akin to that of an effective interfacial tension. Similarly to the theory for immiscible fluids, Korteweg's theory relates the effective surface tension $\Gammae$ to the gradient of composition across the interface. For our colloidal suspensions, one can write
\begin{equation}\label{eq:sq-grad}
\Gammae=  \kappa\int_{-\infty}^{\infty} \left(\frac{d \phieff}{dz}\right)^2 dz \simeq \frac{\kappa}{\delta} \phieff^2 \,,
\end{equation}
where $z$ is the coordinate orthogonal to the interface and $\kappa$ the square-gradient constant. The last approximation holds for a linear concentration profile that increases from 0 (in the solvent) to $\phieff$ (the value in the bulk of the suspensions), over an interface thickness $\delta$.

The dashed line in \fref{fig:Gamma-phi} shows that for the microgels (\ref{eq:sq-grad}) holds over the full range of $\phieff$. This is \textit{a priori} very surprising, since the square-gradient model neglects powers of $d \phieff/dz$ higher than two, which should become important at high $\phieff$. As argued in \cite{Truzzolillo2014}, this apparent inconsistency is removed if one realizes that for the PNiPAM suspensions the relevant parameter is $\varphi$, the volume fraction of the monomers composing the microgels, rather than the effective volume fraction of the microgel particles. Because the microgels are highly swollen, $\phieff/\varphi \approx 27.75$, so that our experiments are actually performed in the low-$\varphi$ limit where (\ref{eq:sq-grad}) is expected to be valid. As a further test of this interpretation, we have calculated in \cite{Truzzolillo2014} the square-gradient constant $\kappa$ by adapting existing polymer theories \cite{Balsara1988}. The interface thickness $\delta$ can then be obtained from $\kappa$ and the prefactor in the quadratic fit to the experimental $\Gammae(\phieff)$ data. The value of $\delta$ thus obtained range from $364$ nm to $392$ nm, comparable to the average interparticle distance \cite{Truzzolillo2014}, as expected for a sharp interface, thus supporting this scenario.

For the Ludox suspension, by contrast, the quadratic law (\ref{eq:sq-grad}) fails to fit the sharp growth of $\Gammae$ \textit{vs} $\phieff$. This is consistent with the above picture, because the Ludox particles are compact objects, such that a sharp composition gradient is established across the solvent-suspension interface. Consequently, higher order terms not included in Korteweg's theory are likely to be important. One would still expect (\ref{eq:sq-grad}) to hold at low enough $\phieff$; unfortunately, the viscosity contrast between diluted Ludox suspensions and their own solvent is too low for this regime to be probed in our Hele-Shaw experiments. In the range of packing fractions of our experiments, we find that the interfacial surface tension of Ludox suspensions is well reproduced by an exponentially diverging law: $\Gammae = a\exp(b\phieff)$, with $a = 1.9$ nN~m and $b=35.8$ (solid line in \fref{fig:Gamma-phi}). Ongoing work in our group focuses on investigating the physical origin of this empirical law, as well as its dependence on particle size, internal structure, and interparticle interactions.

\section{Conclusions}
We have investigated both bulk and interfacial stresses in strongly driven colloidal suspensions, focusing on the analogies and differences between suspensions comprising particles with different internal structure and interparticle interactions. We find that the flow curves of soft, deformable microgels and those of suspensions of hard silica particles are very similar up to packing fractions $\phieff \lesssim \Phi_{rcp}$. This suggests that in our PNiPAM system the fluid-to-solid transition is driven by the same thermal glass transition as for the hard particles, and not by an athermal jamming transition. Only at packing fractions $\phieff > \Phi_{rcp}$ does the deformability of the microgels play a role, as revealed by the onset of a different regime where the yield stress plateau is governed by the energetic cost to deform a particle, rather than by the thermal stress.

In striking contrast to the bulk rheology, the behavior of the (transient) interfacial stress that arises when a sharp concentration gradient is imposed between the suspensions and their own solvent strongly depends on the nature of the particles. The surface tension of microgels can be rationalized in the framework of Korteweg's theory for mild interfacial concentration gradients, because for microgels the monomer concentration $\varphi$ --and thus the interfacial concentration gradient-- remains small even at high $\phieff$. Moreover, a quantitative analysis \cite{Truzzolillo2014} inspired by work on polymer solutions \cite{Balsara1988} allows one to determine the square-gradient constant $\kappa$, and therefore to account for the absolute magnitude of $\Gammae$. This result is based on the calculation of the free energy cost associated to the reduction of the degrees of freedom of the polymers chains near the interface, due to the additional constraint imposed by the presence of a concentration gradient \cite{Balsara1988}. Thus, both the applicability of Korteweg's theory to PNiPAM suspensions and the absolute magnitude of $\Gammae$ are ultimately the consequence of the polymeric nature of the microgels, rather than their deformability \textit{per se}. It would be interesting to further test these ideas by measuring $\Gammae$ for suspensions of soft but non-polymeric particles, \textit{e.g.} using emulsions.

Our understanding of interfacial stresses for hard particles is much less advanced. We find that $\Gammae$ increases exponentially with $\phieff$, as opposed to the quadratic growth seen for the microgels and predicted by Korteweg's theory. Deviations from the square-gradient law are not unexpected for systems exhibiting strong concentration gradients, as is the case for the sharp interface between the solvent and a dense suspension of compact particles. However, the origin of the precise functional form of $\Gammae$, its magnitude, the influence of the particle material, softness, and interparticle interactions are still open question that we are currently investigating.\\

\textit{Acknowledgments}. We acknowledge fruitful discussions with J.-L. Barrat and L. Berthier. This project was supported by ANR under Contract No. ANR-2010-BLAN-0402-1.\\

\textbf{Bibliography}\\

\bibliographystyle{unsrt}
\bibliography{art_JPCM}

\end{document}